\documentstyle[latexsym,psfig]{mn}

\def\h{\hbox{$^{\rm h}$}}
\def\m{\hbox{$^{\rm m}$}}

\def\degr{\hbox{$^\circ$}}

\def\farcm{\hbox{$.\mkern-4mu^\prime$}}
\def\farcs{\hbox{$.\!\!^{\prime\prime}$}}               
\def\sqd{\hbox{$^{\Box}$}}
\def\fsqd{\hbox{$.\!\!^{\Box}$}}

\begin{document} 
\title[The Faint Sky Variability Survey]{The Faint Sky Variability
Survey I: \\ Goals and Data reduction process
}
\author[P.~J.~Groot et al.]{P.J. Groot$^{1,2,3}$, 
P.M. Vreeswijk$^3$,
M.E. Huber$^{4,5}$, 
M.E. Everett$^4$, 
S.B. Howell$^4$, 
\and G. Nelemans$^{3,6}$, 
J. van Paradijs$^{3,7}$, 
E.P.J. van den Heuvel$^3$, 
T. Augusteijn$^{8,9}$
\and E. Kuulkers$^{10}$, 
R.G.M. Rutten$^8$,
J. Storm$^{11}$\\
$^1$Department of Astrophysics, University of Nijmegen, P.O. Box 9010,
6500 GL Nijmegen, The Netherlands\\
$^2$Harvard-Smithsonian Center for Astrophysics, 60 Garden Street,
Cambridge, 02138 MA, USA\\
$^3$Astronomical Institute `Anton Pannekoek'/ CHEAF, Kruislaan
403, 1098 SJ, Amsterdam, The Netherlands\\
$^4$ Astrophysics Group, Planetary Science Institute, 620 N. 6th Ave., Tucson, AZ,  USA\\
$^5$ Department of Physics and Astronomy, University of Wyoming, PO
Box 3905, Laramie, WY 82071, USA\\
$^6$ Institute of Astronomy, University of Cambridge, Madingley Road,
CB3 0HA, Cambridge, UK\\
$^7$ Physics Department, University of Alabama in Huntsville,
Huntsville, USA\\
$^{8}$ Isaac Newton Group of Telescopes, Apartado de Correos 321,
38700 Sta Cruz
de La Palma, Canary Islands, Spain\\
$^{9}$ Nordic Optical Telescope, 
Apartado de Correos 474, 38700 S/C de La Palma, Canary Islands, Spain\\
$^{10}$ ESA-ESTEC, Science Operations \& Data Systems Division, SCI-SDG,
Keplerlaan 1, 2201 AZ Noordwijk, The Netherlands\\
$^{11}$ Astrophysikalisches Institut Potsdam, An der Sternwarte 16,
D-14482, Potsdam, Germany }

\maketitle

\begin{abstract}
The Faint Sky Variability Survey is aimed at finding photometric
and/or astrometric variable objects in the brightness range between
$\sim$16$^{th}$ and $\sim$24$^{th}$ magnitude on timescales between
tens of minutes and years with photometric precisions ranging from 3
millimagnitudes for the brightest to 0.2 magnitudes for the faintest
objects. An area of $\sim$23 square degrees, located at mid and high
Galactic latitudes, has been covered using the Wide Field Camera on
the 2.5m Isaac Newton Telescope on La Palma. Here we describe the main
goals of the Faint Sky Variability Survey and the data reduction
process.
\end{abstract}

\section{Introduction \label{sec:intro}}
The advance of large format ($>$2k$\times$2k) CCDs with high quantum
efficiency has opened up a new area in Galactic and extragalactic
astrophysics: the systematic study of astrophysical objects fainter
than 20th magnitude. The importance of this brightness regime is
nicely illustrated by the current, fast development in the field of
$\gamma$-ray bursts (GRBs; for a recent review see Van Paradijs,
Kouveliotou and Wijers, 2000), where the localization of faint
variable optical counterparts has led to a large increase in our
understanding of GRBs.

In the following sections we will outline the main goals of the Faint
Sky Variability Survey\footnote{http://www.astro.uva.nl/$\sim$fsvs}
(FSVS, Sect.\,2), the
INT Wide Field Camera (Sect.\,3), the observing strategy (Sect.\,4)
and field selection (Sect.\,5).  After a short comparison with other,
running surveys (Sect.\,6), we will discuss data reduction (Sect.\,7),
final data products (Sect.\,8) and availability of the data (Sect.\,9).

\section{Goals of the FSVS \label{sec:goals}}

Understanding the variability of stars 
has often been crucial in the development of
astrophysics, with applications ranging from the evolution of stars, to
the structure of our Galaxy and the distance scale of the Universe. 
Current variability studies are mainly restricted to either bright
regimes (brighter than 20th magnitude) or smaller areas (high-$z$ supernovae
and GRB searches). In the Galaxy, a deep variability study
will not only reveal the characteristics of specific groups of 
stellar objects, but will also shed light on the outer parts of our 
Solar System, the direct Solar Neighbourhood, the structure of our Galaxy,
and the extent of the Galactic halo. The FSVS has observed
$\sim$23 square degrees (23\sqd) down to 24th magnitude. 
The main targets can be divided into two broad areas of 
interest: photometrically and astrometrically
variable objects.

\subsection{Photometrically variable objects \label{sec:photgoals}}
Among the various classes of variable stars our main targets are:
\begin{itemize}
\item {\sl Close Binaries}: \\ Current detections of low-mass
close-binary systems (Cataclysmic Variables, Low-Mass X-Ray Binaries
(LMXBs, including Soft X-Ray Transients, SXTs) and AM CVn stars) are
strongly biased to small subsets of their populations. Of these
systems the Cataclysmic Variables (CVs) form the main subgroup we
expect to find. We refer to Warner (1995a) for an extensive review of
CV properties, Van Paradijs \& McClintock (1995) for optical
observations of LMXBs and Warner (1995b) for AM CVn stars.  Apart
from novae, most CVs are either found as by-products of extragalactic
studies like blue-excess, quasar surveys (e.g. the Palomar-Green
survey: Green, Schmidt and Liebert, 1986; the Hamburg Quasar Survey:
Engels et al., 1994; the Hamburg-ESO Quasar Survey: Wisotzki et al.,
1996; and the Edingburgh-Cape Survey: Stobie et al., 1988), or by
their outbursts in which the system suddenly brightens 3-10 magnitudes
due to enhanced mass transfer through the accretion disk. However,
theoretical calculations show that the majority of the CV population
should have evolved down to mass-transfer rates that are lower than
$\sim$10$^{-11}$ M$_{\odot}$\,yr$^{-1}$ (see e.g. Kolb 1993; Howell,
Rappaport and Politano, 1997; Howell, Nelson and Rappaport, 2001;
however, see Patterson, 2001 for an alternative view).  At these very
low-mass transfer rates, CVs are expected to be faint (typically
V$>$20), have no UV excess, show no (frequent) outbursts, and will
therefore not show up in conventional searches. However, all CVs show
intrinsic variability of the order of tenths of magnitudes or
more. This variability is either caused by `flickering' (mass-transfer
instabilities), orbital modulations (hot spots or eclipses) or
long-term mass-transfer fluctuations. Searching for faint variable
stars is therefore a good way to define the characteristics of the
majority of the CV population.  Based on population synthesis models
we expect to find 20 new CVs per square degree (Howell et al., 1997).
The same search technique will also make the survey sensitive to other
classes of close binaries, such as LMXBs, SXTs in quiescence and AM
CVn stars.
\item {\sl RR Lyrae stars:} \\ Due to their standard candle properties
and easy recognition by colour and variability, RR Lyrae stars can be
used as excellent tracers of the structure of the Galactic halo. A few
of these stars have been found at large galactocentric distances
(Hawkins, 1984; Ciardullo et al., 1989), but number statistics are
still poor. Finding more of these stars will help to constrain the
total enveloped mass in the Galaxy at different radii.  From the small
number of known systems we derive a very uncertain estimate of 0.2 RR
Lyrae stars per square degree that are beyond 30kpc.
\item {\sl Optical Transients to GRBs:}\\ The detection of
optical counterparts to GRBs (e.g. Van Paradijs et al., 1997), and the
subsequent classification of GRBs as cosmological (e.g. Metzger et
al., 1997, Kulkarni et al., 1998) have shown that GRBs are among the
most energetic phenomena known in the Universe. The high energies
implied by observations of GRB afterglows (10$^{53-54}$ erg in
$\gamma$-rays if isotropy is assumed, Kulkarni et al., 1998; 1999),
raises the question whether GRBs are emitting their energy
isotropically or in the form of jets. In the latter case the energies
involved will be much lower, depending on the amount of beaming. Even
if the $\gamma$-rays are beamed the optical afterglow is expected to
radiate more isotropically, and thus one expects to observe faint
afterglows without an accompanying burst in $\gamma$-rays. The
detection rate of such transient events will constrain the beaming
angle. The expected detection rate depends very much on the chosen
geometry of the GRBs and varies for the FSVS database
between several dozens and $\ll$1. A discussion and analysis of our
results is presented in Vreeswijk (2002).
\end{itemize}

\subsection{Astrometrically variable objects \label{sec:astgoals}}
The observing schedule that we have adopted for the FSVS (see
Sect.\,4) also allows for the detection of astrometrically variable
objects. Our interests fall into two main categories: 
\begin{itemize}

\item {\sl Kuiper Belt Objects}:\\ Kuiper Belt Objects (KBOs) are icy
bodies revolving around the Sun in orbits that lie outside the orbit
of Neptune (which has led to the alternative name of Trans Neptunian
Objects; TNOs). Since their discovery in 1993 (Jewitt and Luu, 1993),
more than 100 of these objects have been found. Studying their
properties will give important insight into the formation of the Solar
system and planetary systems in general.  One question that is
particularly well suited to be answered is the inclination
distribution of KBOs. Most KBOs have been found within 5\degr\ from
the ecliptic, but this may constitute an observational bias, since
most searches have been (and are) performed close to the
ecliptic. Since the FSVS is mostly pointing away from the ecliptic, we
will be able to set limits on the inclination distribution of KBO's.
KBO's are found from intra-night astrometric variability. 
One KBO is found so far in a preliminary search of the FSVS database
(K01QW2X; Gladman et al., 2001).

\item {\sl Solar Neighbourhood Objects}: \\The yearly re-observations
allow for the detection of high proper-motion objects in the Solar
neighbourhood. These will be extremely important to constrain the
low-mass end of the IMF in the solar neighbourhood, to estimate the
relative contribution of the disk and halo population of stars in the
solar neighbourhood and trace the star formation history of the
Galactic halo by finding old, high proper motion, white dwarfs.
\end{itemize}

\section{The INT Wide Field Camera \label{sec:wfc}}  
The Wide Field Camera\footnote{see:
http://www.ast.cam.ac.uk/$\sim$wfcsur/index.php for an extensive
description of the WFC} (WFC) is mounted at the prime focus of the
2.5m Isaac Newton Telescope (INT) at the Observatorio del Roque de Los
Muchachos on the island of La Palma. The WFC consists of 4 EEV42 CCDs,
each containing 2048$\times$4100 pixels.  They are fitted in an
L-shaped pattern, which makes the Camera 6k$\times$6k, minus a
2k$\times$2k corner (see Figure 1).  The CCDs consist of 13.5$\mu$
pixels (0\farcs33 per pixel on the sky), which gives a sky coverage
per CCD of 22\farcm8$\times$11\farcm4. A total of 0\fsqd29 is covered
by the combined four CCDs.  With a typical seeing of 1\farcs0-1\farcs3
on the INT, point objects are well-sampled, which allows for accurate
photometry. The Camera is equiped with Harris and Sloan filters, of
which we use the Harris $B, V$ and $I$ filter. Zeropoints, defined as
the magnitude that gives 1 detected e$^-$/s, in these filters are 25.6
in $B$ and $V$ and 25.0 in $I$.

\begin{figure}
\centerline{\psfig{figure=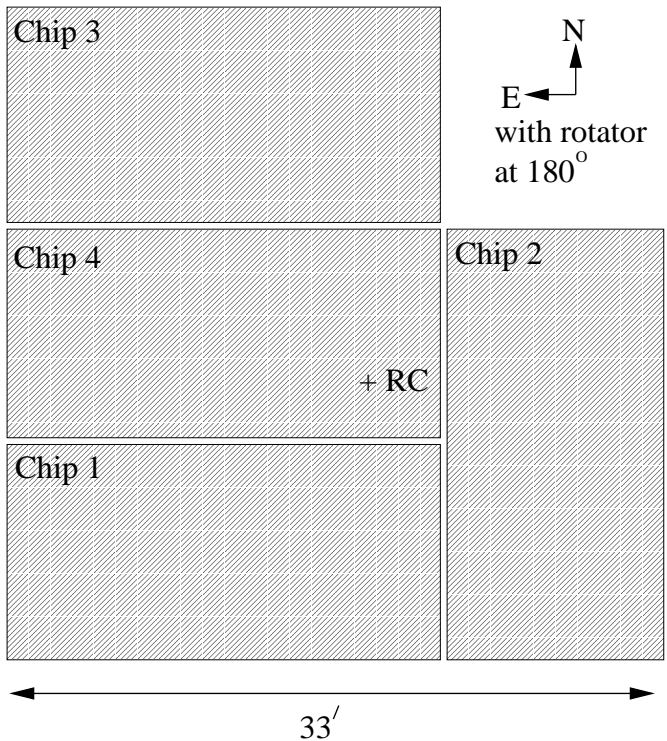,width=8.8cm,angle=-90}}
\caption[]{Graphical lay-out of the WFC 4 EEV 4k$\times$2k CCDs. In
the orientation used by the FSVS, North is up and East is to the
left. The WFC rotates around its Rotator Center (RC). \label{fig:layout}}
\end{figure}

\section{Observing strategy \label{sec:strategy}}

The typical timescales of variability covered by the objects listed
above vary from hours (CVs, KBOs, RR Lyrae stars) to days (optical
transients to GRBs) to years (high proper motion stars). To cover all
possible timescales of variation we have devised an observing strategy
that optimises both the coverage per field as well as the total sky
coverage. The variability search is done with 10 min. V-band
observations. This is a compromise between the expected colours of our
targets and the sensitivity of the WFC which peaks between 4000\AA\
and 6000\AA\ or the photometric variability we find that at least
15-20 pointings are needed to firmly state that an object is variable
and also get an indication of the timescale of its variability (or
ideally its period). For the first two runs of the FSVS this number
was limited to $\sim$10, but has been raised to 15-20 in subsequent
runs.

The FSVS has been observing in one-week time slots, separated by
roughly half year intervals. Observations of one field are mainly 
obtained within the one week observing run, with single observations
in the yearly returns. In the observing sequences
we have tried to avoid a regular spacing of the observations since
this will introduce strong aliases in any period search. On
photometric nights (which were always present in each of the runs) the
fields were observed in B (10 min) and I (15 min) together with
Landolt (1992) fields. These observations were taken
centered on each of the four chips to obtain a sufficient number of
photometric standards per chip (see Sect.\ \,7.9). 
Using this observing strategy an average of 4\sqd per one-week run was
observed. Single V-band re-observations of each field are being
obtained on a yearly basis.

\section{Field selection \label{sec:selection}}

The field selection was governed by the following four criteria (in
order of importance) to ensure maximum quality of the data:
\begin{itemize}
\item Fields are located at Galactic latitudes $b^{II} >$ 20\degr: to
probe the Galactic halo as well as the Galactic disk to considerable
depths we target most of our fields at mid-Galactic latitudes (see
Table\ \ref{tab:fields}). This also prevents problems with field
crowding and interstellar extinction that will be present at lower
Galactic latitudes. 
\item Fields are observed within a zenith distance, $z<$30\degr:
this criterion has been set to limit the effect of differential
extinction coefficients on the accuracy of the photometry.
\item If possible we select our fields at the ecliptic, to
increase the chances of finding KBOs. However, as explained in Sect.\
\ref{sec:astgoals} even if we are not able to point at the ecliptic,
our results may help to constrain the inclination distribution of
KBOs.
\item Bright stars are avoided: stars brighter than $\sim$10th
magnitude will cause large charge overflows and diffraction patterns
that limit the area on a CCD that can be used for accurate
photometry, depending on the placement and brightness of the star. To
prevent this from happening the fields are selected to be as devoid as
possible of bright stars. We checked for the presence of
bright stars using the Digital Sky Survey  in the selection of the fields. 
\end{itemize}
It is clear that not all four criteria can always be met. 
For the Northern Hemisphere all four criteria can  only
be met in late November-early December. Table 1 shows the center
points of the FSVS fields, together with the Galactic 
coordinates and period of first observations. 

\begin{table}
\caption{Field centers and period of observations of the FSVS
fields. All coordinates are in J2000 units.\label{tab:fields}}
\begin{tabular}{lllrrl}
Field No. & RA & Dec & $l^{\sc ii}$ & $b^{\sc ii}$ & Period\\
1-6 	& 23\h44\m & +27\degr15\arcmin & 105 & --33 & Nov 98\\
7-12	& 02\h32\m & +15\degr00\arcmin & 156 & --40 & Nov 98\\
13-18	& 07\h52\m & +20\degr40\arcmin & 200 & +22 & Nov 98\\
19-22   & 12\h53\m & +27\degr01\arcmin & 220-21 & +90 & May 99\\
23-26	& 12\h51\m & +26\degr20\arcmin & 268-360 & +89 & May 99\\
27-30 	& 16\h25\m & +26\degr33\arcmin & 45 & +43 & May 99\\
31-34	& 17\h20\m & +27\degr00\arcmin & 49 & +31 & May 99/00\\
35-40 	& 03\h02\m & +18\degr38\arcmin & 161 & --33 & Jan 00/01\\
41-46,59 	& 07\h15\m & +21\degr00\arcmin & 196 & +15 & Jan 00/01\\
47-52,60	& 10\h00\m & +21\degr30\arcmin & 211 & +50 & Jan 00/01\\
52-56	& 16\h23\m & +27\degr03\arcmin & 45 & +42 & May 00\\
57-58	& 16\h32\m & +21\degr16\arcmin & 39 & +39 & May 00\\  
61-62	& 10\h37\m & +04\degr00\arcmin &242 & +50 & Jan 01\\
63-66   & 17\h25\m & +27\degr30\arcmin & 50 & +30 & Jul 01\\
68-71   & 22\h02\m & +27\degr30\arcmin & 83 & --21 & Jul 01\\
72-75   & 18\h32\m & +36\degr00\arcmin & 64 & +19  & Aug 01\\
76-79 	& 23\h47\m & +28\degr10\arcmin &106 & --32 & Aug 01
\end{tabular}
\end{table}

\section{Comparison with other surveys \label{sec:comparison}}

The FSVS is unique in its search for variability on short timescales
(tens of minutes to days), depth and precision of its differential
photometry, although having a rather moderate sky-coverage. The Sloan
Digital Sky Survey (SDSS; York et al., 2000) covers a much larger area
of the sky (10\,000 \sqd), but at brighter magnitudes (14
$<g^{\prime}<$ 22.5), and provides almost no variability
information. The microlensing studies (e.g. MACHO, Alcock et al.,
1997; EROS, Beaulieu et al., 1995; OGLE, Udalski et al., 1992) do
obtain variability information, but are targeted at different stellar
populations (the Galactic Bulge, the LMC, or M31) and have a limit of
V$\sim$21 with a photometric precision of 0.5 mag at the faint end,
caused by limited S/N and crowding in their necessarily high density
star fields. High $z$ supernovae searches reach as deep as the FSVS,
but have a lower time-resolution and smaller area.  In Figure 2 we
show schematically how the FSVS compares with other deep ongoing
surveys.

\begin{figure}
\centerline{\psfig{figure=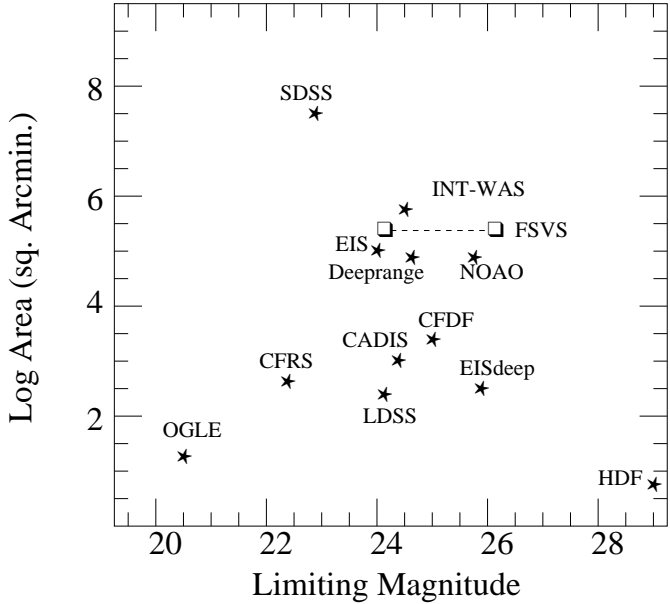,angle=-90,width=8.8cm}}
\caption[]{A comparison in area and depth between major current
surveys and the FSVS. Adapted from the NOAO Deep Survey Web-pages (see
http://www.noao.edu/noao/noaodeep/; SDSS=Sloan Digital Sky Survey,
York, et al., 2000; EIS(deep) =
ESO Imaging Survey (Deep), Nonino et al., 1999; Deeprange = Postman et
al. 1998; INT-WAS: INT Wide Angle Survey: McMahon et al., 2001; HDF=
Hubble Deep Field, Williams et al., 1996; NOAO= NOAO Survey, Jannuzi
and Dey, 1999; CFRS = Canada France Redshift Survey, Lilly et al.,
1995; CADIS = Calar Alto Deep Imaging Survey, Hippelein et al., 1998;
CFDF = Canadian French Deep Fields; Brodwin et al., 1999;
LDSS=Glazebrook et al., 1995; OGLE = Optical Gravitational Lensing
Experiment, Udalski et al., 1992 ).  Note
that most of these surveys have no or very limited variability
information. The range in depth for the FSVS corresponds to using each
individual image (as in the variability study) or the sum images.}
\end{figure}

\section{Reduction and Analysis Methods \label{sec:reduction}}

To obtain variability information on all the objects detected in our
observations we use the technique of differential aperture and psf
photometry. We have written a pipe-line reduction package, consisting
of IRAF tasks, Fortran programs and at its core the SExtractor program
by Bertin and Arnouts (1996). 
Every object in every observation is \mbox{analysed} and the results
are stored in a master-table that lists the essential information
(described below in detail) for each object. Below we outline the data flow
through our pipe-line reduction, starting with the raw data as it
comes from the telescope. 

\subsection{Bias subtraction \label{sec:bias}} 

The mean of the counts in the overscan region of each observation is
used to subtract the overall bias level. After this the 2-D bias
pattern, determined from bias observations taken at the start of the
night, is subtracted.

\addtocounter{footnote}{-1}
\subsection{Linearization of the data \label{sec:linear}}

A non-linearity in the read-out electronics causes all data taken with
the INT WFC to be non-linear up to a level of $\sim$5\%.  The
magnitude of this non-linearity as a function of exposure level is
determined by the Cambridge WFS group\footnote{ see: previous footnote
for URL and details} and is posted in tabular and analytic form. These
corrections are applied after bias-subtraction.

\subsection{Flatfielding \label{sec:flatfield}}

From twilight skyflats taken during a complete observing run a
master flatfield is made, which is used for all the observations taken in
that band during the observing run. For the I-band observations, which
suffer from fringing at the 3.5\% level, we have made fringe maps from
the night time observations, which allows the fringe pattern to be removed
down to the 0.6\% continuum sky level (see Fig.\ \ref{fig:fringe}).

\begin{figure*}
\begin{minipage}{5cm}
\end{minipage}
\begin{minipage}{5cm}
\end{minipage}
\begin{minipage}{5cm}
\end{minipage}
\caption[]{Defringing of the I-band observations, using a fringe map
made from the night-time observations themselves. Left: Before
defringing, middle: fringe map, right: after defringing \label{fig:fringe}}
\end{figure*}

\subsection{Source detection \label{sec:detection}}

The bias-subtracted, linearized and flat-fielded data are fed to the
SExtractor program. This program detects sources and measures their
instrumental magnitude in a number of different ways, as set by the
user. Source detection is done by requiring that three neighbouring
pixels are more than two sigma above the sky-background. Visual
inspection shows that this threshold value is capable of detecting
virtually all objects that can be identified by eye. Some
contamination from extended cosmic rays is present, but these are
effectively removed in the subsequent steps. Apart from finding the
sources and determining their instrumental magnitudes, for each source
the SExtractor program determines other characteristic parameters
such as the position, size, extent, ellipticity and orientation angle.
Due to vignetting a corner of CCD3 (the NE corner in Fig.\
\ref{fig:layout}) has very low count rates. We discard any object
detected in a square box 200 pixels wide from this corner of CCD3.
Spatial offsets between observations of the same fields in different
visits are small (typically $<$20\arcsec) so no check is made for
detections on different CCDs. 

\subsection{Instrumental magnitudes\label{sec:instrumental}}

\begin{figure}
\centerline{\psfig{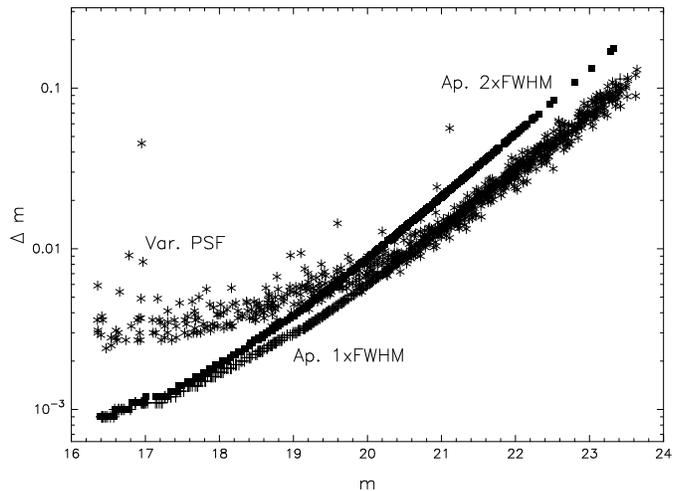}}
\caption[]{Instrumental photometric errors per magnitude for seeing
matched aperture photometry with 1$\times$FWHM (+-signs),
2$\times$seeing FWHM (filled squares) and variable psf fitting
(stars). \label{fig:errorpsf} }
\end{figure}

For each object instrumental magnitudes are extracted in four
different ways: fixed aperture photometry, seeing matched aperture
photometry, variable psf fitting photometry and isophotal
magnitudes. The isophotal magnitudes and fixed 12 pixel radius aperture
photometry have been included as a check on the others and for
extragalactic work in case of the isophotal magnitudes.  The seeing
matched aperture photometry uses an aperture that scales with the
seeing of the observations. This seeing is determined using bright
unsaturated stars in the inner 1k $\times$ 1k region of each chip. The
size of the 
aperture for the observation is set to twice the FWHM of the seeing
estimate. This relatively large aperture works well for bright stars,
but the S/N deteriorates for faint objects due to the dominant sky
background. In Fig.\ \ref{fig:errorpsf} we show the variation of the
error for an aperture of twice the seeing FWHM and for an aperture of
one time the FWHM. This clearly shows that for bright stars they work
equally well and for faint stars small apertures work better.
However, in Sect.\ \ref{sec:difmags} we will show that the
 aperture photometry is not ideal for variability studies. 

The error on the instrumental magnitudes is determined only by
Poisson-statistics, including the source and background brightness,
read-out noise of the chip and the gain.

In the variable psf fitting the point-spread-function of the objects,
and its variation, over the chip is determined from a set of 25
isolated stars, spread equally in position over the chip. Using this
variable psf the instrumental magnitude of each object is determined
with the use of the IRAF {\sc daophot} package, in which the
photometric accuracy is adjusted from 1 mmag to 0.1 mmag to be able to
obtain the photometric precision needed for the brightest stars.  The
error in the instrumental magnitude is now a combination of the
Poisson statistics (as in the aperture photometry) as well as a factor
from the fitting procedure. This causes the psf errors on the
brightest objects to be higher than in the aperture photometry (see
Fig.\ \ref{fig:errorpsf}), but for the faint sources it is as good as
that of the 1$\times$FWHM aperture photometry.

Although from just Fig.\ \ref{fig:errorpsf} it would seem that the
aperture photometry is doing much better than the psf photometry at
the brighter end, the variability study (as discussed in Sect.\
\ref{sec:difmags}) shows this not to be the case. The Poisson errors
on the aperture photometry are not a good representation of the actual
error on the measurement. Other sources of errors besides counting
statistics such as low-level gain and read-out noise variations,
flatfield errors and a variable psf become dominant at bright
magnitudes. This shows up as a flattening of the error
distribution. In Sect.\ \ref{sec:difmags} we will show that the small
aperture photometry errors introduce apparent variability for the
brightest stars and therfore aperture photometry is therefore not the
most suitable for the FSVS.
 
\subsection{Field matching \label{sec:matching}}

Different observations of the same field are automatically matched
using the {\sc offset} program, supplied with the {\sc dophot} package
(Schechter, Mateo and Saha; 1993), using the 100 brightest,
non-saturated stars, that are not located near the edges of the
CCDs. Matching is done by triangle pattern recognition in the two
images. This matching allows for linear scaling, rotation and
translation of the different images. Output is given as the elements
of a rotation-translation matrix. All image source catalogues are
transformed to that of a reference image (the one with the best
seeing). Individual objects are matched if in the new image an object
is found within 1 FWHM of the position of the object in the reference
image.  This same criterion is used to match stars between different
filters.

\subsection{Local reference star selection \label{sec:standards}}

In order to obtain differential magnitudes, an ensemble of local
reference stars has to be selected. The average (ensemble) magnitude
of these stars is used as a baseline to compute all instrumental
magnitudes. In the selection of this ensemble it is important to use
the brightest, non-variable, stars that are not saturated. Using the
brightest stars is essential because the error on the differential
magnitude of any object consists of the error that is obtained from
counting statistics for that object, and the error on the average of
the reference stars (see e.g. Howell, Mitchell and Warnock, 1988). The
uncertainty in the mean magnitude of the ensemble must be made
significantly smaller than the uncertainty imposed by counting
statistics on the magnitude of any star of interest.  If this is not
the case, it will cause small-amplitude variability, that should have
been detected on the basis of counting statistics, to become
undetectable. Per CCD, an ensemble of ten local standards is selected
by requiring that their variation with respect to the average is less
than 5 millimagnitudes. If this requirement is set more stringently
not enough standards are found. In the Galactic North Pole
observations of May 1999 the selection criterion had to be relaxed to
10 millimagnitudes in order to find a suitable number of stars. This
is, of course, due to the limited number of stars in the NGP
direction. As explained above, this selection criterion naturally sets
the minimum amplitude (= scatter/$\sqrt{N_{\rm reference\ stars}}$) of
variation that can be found. We have not taken a colour difference
between the ensemble and targets stars into account. However, this
small effect will only be important for the brightest stars, which, on
average, will also have similar colours to the ensemble stars. 

\subsection{Differential magnitudes and variability\label{sec:difmags}}

\begin{figure}
\centerline{\psfig{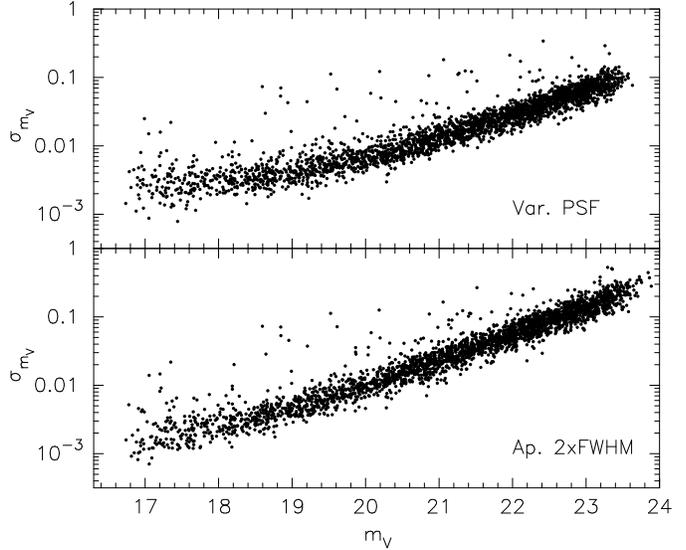}}
\caption[]{The standard deviation on the light curves of point sources
(stellarity $>$0.8) in the same field as shown in Fig.\
\ref{fig:errorpsf}. {\sl Top} for Variable PSF fitting and {\sl
bottom} for 2$\times$seeing FWHM aperture photometry. 
\label{fig:sigm}}
\end{figure}

For every object the differential magnitude is calculated against the
ensemble average. The error of the instrumental magnitude is
propagated to the differential magnitude, adding quadratically to the
error on the ensemble average. The error on the ensemble average is
determined from the scatter of the ensemble stars at that epoch around
their average over all epochs. The differential magnitude is
calculated for all four instrumental magnitudes as described in Sect.\
\ref{sec:instrumental} for all observations of this field. In Fig.\
\ref{fig:sigm} we show the variation around their average magnitude
for all objects in a representative field of the FSVS, both for seeing
matched aperture photometry as well as for the variable PSF
fitting. The rise towards fainter magnitudes is a consequence of the
larger instrumental magnitudes errors due to lower count rates.  For
the brightest sources a differential magnitude variation of $<5$mmag,
which is at the level of extrasolar planet transits, is easily
obtained.

\begin{figure}
\centerline{\psfig{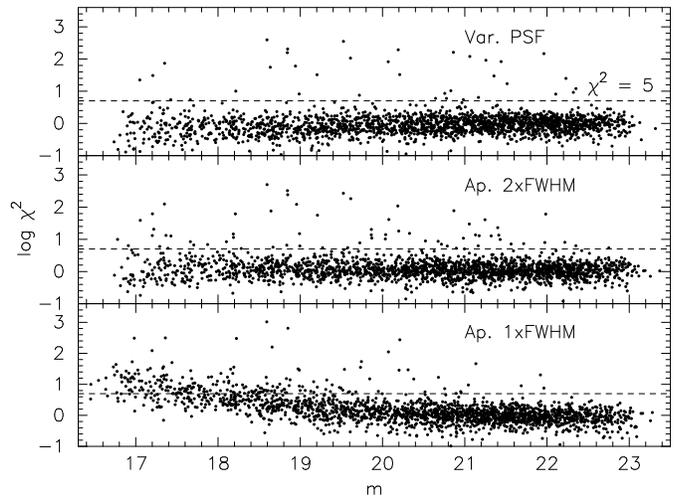}}
\caption[]{Variability distributions for the same field as shown in
Figs.\ \ref{fig:errorpsf} and \ref{fig:sigm}. {\sl Bottom:} seeing
matched aperture photometry with aperture size equal to 1$\times$
seeing FWHM. {\sl Middle:} same as bottom but with aperture size
2$\times$FWHM. {\sl Top:} variable psf fitting. \label{fig:chi2} }
\end{figure}

Variability is determined by calculating the reduced $\chi^2$ value of
the light curve with respect to its average value. As expected this is
a constant as a function of magnitude (Fig.\ \ref{fig:chi2}). In Fig.\
\ref{fig:chi2} we show the $\chi^2$ distribution for the 1$\times$FWHM
aperture (bottom), and 2$\times$FWHM aperture photometry (middle) and
the variable psf fitting (top). The dashed line in Fig.\
\ref{fig:chi2} shows the 5-$\sigma$ variability level above which we
denote our stars to be variable. From this we see that an aperture of
1$\times$FWHM is too small for the bright stars and introduces
spurious variability. The 2$\times$FWHM also suffers from spurious
variability, although that is not immediately clear from Fig.\
\ref{fig:chi2}. 

Despite the accurate photometry on a single epoch, the 2$\times$FWHM
aperture photometry suffers from the introduction of systematic
variability into the light curves due to the basic assumption of
aperture photometry that the psf is the same for all objects in the
field. The chips of the WFC are slightly tilted with respect to the
focal plane of the camera, which introduces a variation in the psf of
$\sim$20\% over the field of a single chip. When analysed with
aperture photometry this introduces spurious variability both at the
bright end as well as at the fainter end of the magnitude range. At
the bright end the variation is caused by the change of the psf due to
tilt of the CCDs. At the fainter end the change is caused by barely
resolved binaries and compact galaxies. Not including this source of
error in the aperture photometry at the bright end causes the high
number of spurious variables. In the PSF fitting these errors are
taken into account (as can be seen from the higher level of
single-epoch errors in Fig.\ \ref{fig:errorpsf}), and the spurious
variability is removed (see Fig.\ \ref{fig:sigm} and \ref{fig:chi2}).
The fact that the average $\chi^2$-value for the psf-fitting lies
around 1 for the non-variable stars at all magnitudes shows that the
psf errors are an accurate reflection of the true uncertainties on the
individual photometric measurements.

\subsection{Astrometric and photometric calibration \label{sec:abscal}}

Using the USNO A2.0 catalogue an astrometric solution is
obtained for each CCD and each field separately. 
On average, we use 20-30 USNO A2.0 stars, which 
is sufficient to obtain a cubic solution that is accurate to
0\farcs2-0\farcs4 in right ascension and declination, depending on the
position of a field on the sky.

\addtocounter{footnote}{-1}
During each of our runs, we have had photometric nights, during
which all fields and several Selected Areas of Landolt (1992) were
observed. After having found the astrometric solution for the
standard stars, we can measure the standard stars automatically.  We
use the SExtractor aperture photometry option, with an aperture radius
of twice the image FWHM. For each CCD the measured $B,V$ and $I$
standard star magnitudes are fitted to a model that includes a
zero-point offset, an airmass term and a colour term. When sufficient
standards are observed at different airmasses, we fit for the airmass
term. If not, we hold it constant at the following values: 0.25, 0.15
and 0.07 for the filters B, V and I, respectively\footnote{see
first WFC footnote}.  The colour term is only
included if it improves the fit significantly.  These solutions are
applied to all objects listed in the catalogue through the ensemble
reference stars that are selected for each CCD of each field (see
Sect. 7.7). From the scatter in the solutions, we estimate the error
in the absolute calibration to be 0.05 for the B and V filters, and
0.1 for the I band. 




\subsection{Limiting magnitudes \label{sec:limmag}}

Based on the amount of flux in the ten
reference stars (see Sect.\ \ref{sec:standards}), the level of the
background sky, the photometry aperture size and the background
aperture size, we calculate the a 3-, 5-, and 7-sigma limiting
magnitude object for each CCD, field and observation. 
In this calculation we neglect the read-out noise since our
observations are long and have background levels whose noise
is much higher than the read-out noise. On average the 5-sigma
limiting magnitudes range between 22.5-24.5 for the B and V-band
images (depending on seeing and cloud cover) and between  21.5 and
23.5 for the I-band observations.



\subsection{Star - Galaxy separation}

\begin{figure}
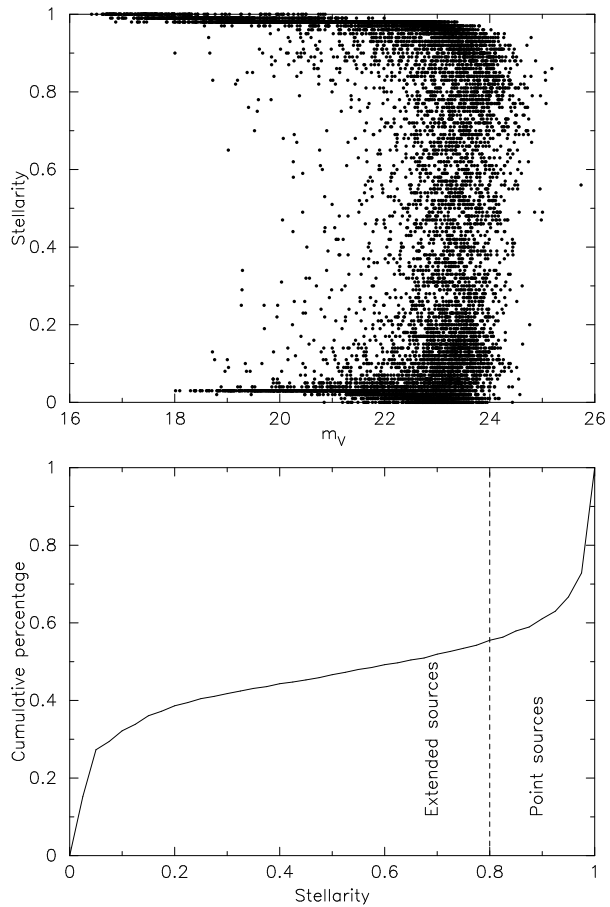

\centerline{\psfig{figure=stellarity.ps,width=8cm,angle=-90}}
\centerline{\psfig{figure=stellarity2.ps,width=8cm,angle=-90}}
\caption[]{{\sl Top:} The stellarity versus magnitude for one of our
fields. A stellarity of zero indicates a highly extended source, and a
stellarity of one is a point-source.  Detections at V$>$25 are noise
spikes. {\sl Bottom:} The cumulative distribution of sources over
stellarity values down to V=24. Using a point source cut-off of 0.8, we have
$\sim$45\% of objects as point sources.
\label{fig:stellarity}}
\end{figure}

The star-galaxy separation used in the FSVS is based on the
'stellarity' parameter, as returned from the SExtractor routines
(Bertin and Arnouts, 1996).  This parameter has a value between 0
(highly extended) and 1 (point source). In the FSVS the stellarity 
value of an object is taken as the value in the combined V-band
images. Due to the increased S/N in this image, the star-galaxy
separation can be done reliably almost 1 magnitude deeper than from 
any individual image. As can be seen in Fig.\
\ref{fig:stellarity} this separation of object types works very well
to classify point-sources (with a value $>$0.8) down to V$\sim$23.5-24.
Fainter stars tend to have slightly lower stellarity values (they turn
down between V=23 and V=24) but can still be well separated from the
galaxies, although some stars at the faint end of the distribution
may be mis-classified as extended. 

\subsection{Astrometric variables}

The proper motion analysis is currently not included in the standard
pipe-line reduction but is handled separately using either the reduced
images (in the case of Kuiper Belt Objects) or the SExtractor output
and astrometric solution as provided by the pipeline (in the case of
the high proper motion stars). Details on both analyses will be given
in subsequent papers. Our first results indicate that proper motions
will be measurable down to 10 mas yr$^{-1}$ with a 2-year baseline. 

 


\section{Final products \label{sec:products}}

The pipeline discussed above returns two sets of output files:\\
$\bullet$ The reduced images\\
$\bullet$ The data tables with the photometric and astrometric
information.\\
The data tables are made per field, per CCD and are made for four
different magnitudes: the psf magnitude, the fixed aperture magnitude,
the isophotal magnitude and seeing matched aperture magnitude.

The header of the table contains all relevant information on the
exposures: run numbers of the original frames, the HJDs of the
observations, the filter, the airmass, the average FWHM of the point
sources in the observation (the 'seeing'), the six element
rotation-translation matrix to the reference image and the 3-, 5-, and
7-$\sigma$ limiting magnitude of the image.
The data tables contain, for each detected object: name, position and
colour, followed by the magnitude, error on the
magnitude, fwhm, stellarity and the error flag as returned from the
SExtractor program for each observation. 

If an object is only detected in a subset of all the observations, it
is added to the final catalogue, and dummy values (any decimal
combination of 9's, e.g. 99.999 , 9.99 etc.) are introduced when it
was not detected.

The object names are given in standard IAU format as
FSVSJhhmmss.ss+ddmmss.s, all in J2000 coordinates. Each object is also
given an 'internal' name whose format is F\_XX\_Y\_ZZZZZ, with XX the
field number, Y the CCD number (1-4) and ZZZZZ a five digit detection
number. The position of each object is given both in RA and DEC as
well as in x,y-coordinates in the reference frame of the specific
field.
  

\section{Availability of the data \label{sec:data}}

All raw images are available upon request from the ING-WFS archive in
Cambridge after the one year propietary right. For UK and NL
astronomers the data is immediately available.
All ASCII data-tables, containing the reduced information described
above, are retrievable from the
FSVS website.



\section{Conclusions}

The FSVS offers a unique possibility of studying the behaviour
of variable objects in the magnitude range of 16 $< V <$ 24 with
photometric precisions ranging from 3 millimag (at $V$=16) to 0.2 mag (at
V$\simeq$24). 

Besides the study of variable objects, the FSVS offers a large dataset
that can serve as the basis for many research topics (e.g. YSO's,
gravitational lenses, galaxy counts, quasar searches). The
FSVS-collaboration encourages the use of the data set for all purposes.


\section*{acknowledgement}

PJG, PMV, GN and the Faint Sky Variability Survey are supported by NWO
Spinoza grant 08-0 to E.P.J.van den Heuvel. PJG is also supported by a CfA
Fellowship. SBH acknowledges partial support of this research from NSF
grant AST 98-19770. MEH is partially supported by a NASA/Space Grant
Fellowship, NASA Grant \#NGT-40008. The FSVS is part of the INT Wide
Field Survey. The INT is operated on the island
of La Palma by the Isaac Newton Group in the Spanish Observatorio del
Roque de los Muchachos of the Inst\'{\i}tuto de Astrof\'{\i}sica de Canarias.


\begin{thebibliography}{}
\bibitem[1997]{macho:97}
Alcock, C., et al., 1997, ApJ 486, 697
\bibitem[1995]{eros:95}
Beaulieu, J.P., et al., 1995, A\&A 303, 137
\bibitem[1996]{bertin:96}
Bertin, E. and Arnouts, S., 1996, A\&AS 117, 393
\bibitem[1999]{bridwin:99}
Brodwin, M., Lilly, S. and Crampton, D., 1999, in {\sl Photometric
Redshifts and the Detection of High Redshift Galaxies}, ed., Weymann
et al., ASP Conf. Ser. 191, p. 105
\bibitem[1989]{ciardullo:89}
Ciardullo, R., Jacoby, G.H., Bond, H.E., 1989, AJ 98, 1648
\bibitem[1994]{HS:94}
Engels, D., Cordis, L., K\"ohler, T., 1994, in {\it IAU Symp. 161},
eds. H.T. MacGillivray et al. (Kluwer, Dordrecht), p. 317
\bibitem[2002]{gladman:02}
Gladman, B.; Kavelaars, J.; Wasserman, L. H., et al., 2001, MPEC 2001-V57
\bibitem[1995]{glazebrook:95}
Glazebrook, K., Ellis, R., Colless, M., et al., 1995, MNRAS 273, 157
\bibitem[1986]{pg:86}
Green, R.F., Schmidt, M. and Liebert, J., 1986, ApJS, 61, 305
\bibitem[1984]{hawkins:84}
Hawkins, M.R.S., 1984, MNRAS 206, 433
\bibitem[1998]{hippelein:98}
Hippelein, H., Beckwith, S., Fockenbrock, K., et al., 1998, in {\sl New
Horizons from Multi-Wavelength Sky Surveys}, ed. McLean,
Golombek, Hayes, and Payne, IAUS 179, p 292
\bibitem[1988]{steve:88}
Howell, S.B., Mitchell, K.J and Warnock, A., 1988, AJ 95, 247
\bibitem[1997]{steve:97}
Howell, S.B., Rappaport, S. and Politano, M.R., 1997, MNRAS 287, 929
\bibitem[2001]{steve:01}
Howell, S.B., Nelson, L. and Rappaport, S., 2001, ApJ 550, 897
\bibitem[1999]{januzi:99} Jannuzi, B.T. and Dey, A., 1999, in {\sl Proc. Photometric redshifts and the Detection of High Redshift
Galaxies}, ed. Weymann et al., ASP Conf. Ser. 191, p. 111
\bibitem[1993]{luu:93}
Jewitt, D. and Luu, J., 1993, Nature 362, 730
\bibitem[1993]{kolb:93}
Kolb, U., 1993, A\&A 271, 149
\bibitem[1998]{shri:98}
Kulkarni, S.R., et al., 1998, Nature 393, 35
\bibitem[1999]{shri:99}
Kulkarni, S.R., et al., 1999, Nature 398, 389
\bibitem[1992]{landolt:92}
Landolt, A.U., 1992, AJ 104, 340
\bibitem[1995]{lilly:95}
Lilly, S.J., Le Fevre, O., Crampton, D., Hammer, F. and Tresse, L.,
1995, ApJ 455, 50
\bibitem[2001]{mcmahon:01}
McMahon, R., et al., 2001, New Astr. Rev. 45, 97
\bibitem[1997]{metz:97}
Metzger, M.,R., et al., 1997, Nature 387, 878
\bibitem[1999]{nonino:99}
Nonino, M., et al., 1999, A\&AS 137, 51
\bibitem[2001]{patterson:01}
Patterson, J., 2001, PASP 113, 736
\bibitem[1998]{postman:98}
Postman, M., Lauer, T.R., Szapudi, I., Oegerle, W., 1998, ApJ 506, 33
\bibitem[1993]{dophot:93}
Schechter, P.L., Mateo, M. and Saha, A., 1993, PASP 105,1342
\bibitem[1988]{ec:88}
Stobie, R.S., Morgan, D.H., Bhathia, R.K., Kilkenny, D. and
O'Donoghue, D., 1988, in {\it The Second Conference on Faint Blue
Stars}, IAU Colloq. 95, eds. D. Philip et al., David Press, Schenectady,
NY, p. 43
\bibitem[1992]{udalski:92}
Udalski, A., et al., 1992, Acta Astronomica 42, 253
\bibitem[1997]{jvp:97}
Van Paradijs, J., Groot, P.J., Galama, T.J et al., 1997, Nature 386, 686
\bibitem[2000]{jvp:2000}
Van Paradijs, J., Kouveliotou, Ch. and Wijers, R.A.M.J., 2000, ARA\&A
38, 379
\bibitem[2002]{pmv:02}
Vreeswijk, P.M., 2002, PhD Thesis, University of Amsterdam
\bibitem[1995]{warner:95}
Warner, B., 1995a, {\it Cataclysmic Variables}, Cambridge Astrophysics
Series 28, CUP, Cambridge, UK.
\bibitem[1995]{warner:95b}
Warner, B., 1995b, Ap\&SS 225, 249 
\bibitem[1996]{HDF:96}
Williams, R., et al., 1996, AJ 112, 1335
\bibitem[1996]{HE:96}
Wisotzki, L., K\"ohler, T., Groote, D and Reimers, D., 1996, A\&AS
115, 227
\bibitem[2000]{sdss:00}
York, D.G., et al., 2000, AJ 120, 1579
\end{thebibliography}
\end{document}